\documentstyle[twoside]{article}

\catcode`\@=11
\long\def\@makefntext#1{
\protect\noindent \hbox to 3.2pt {\hskip-.9pt
$^{{\eightrm\@thefnmark}}$\hfil}#1\hfill}               

\def\thefootnote{\fnsymbol{footnote}}
\def\@makefnmark{\hbox to 0pt{$^{\@thefnmark}$\hss}}    

\def\ps@myheadings{\let\@mkboth\@gobbletwo
\def\@oddhead{\hbox{}
\rightmark\hfil\eightrm\thepage}
\def\@oddfoot{}\def\@evenhead{\eightrm\thepage\hfil
\leftmark\hbox{}}\def\@evenfoot{}
\def\sectionmark##1{}\def\subsectionmark##1{}}

\oddsidemargin=\evensidemargin
\renewcommand{\thefootnote}{\fnsymbol{footnote}}

\newcommand{\Pp}[3]{#1.#2\!\times\!10^{-#3}}

\newcommand{\ba}{\begin{eqnarray}}
\newcommand{\ea}{\end{eqnarray}}

\newcounter{sectionc}\newcounter{subsectionc}\newcounter{subsubsectionc}
\renewcommand{\section}[1] {\vspace{12pt}\addtocounter{sectionc}{1}
\setcounter{subsectionc}{0}\setcounter{subsubsectionc}{0}\noindent
        {\tenbf\thesectionc. #1}\par\vspace{5pt}}
\renewcommand{\subsection}[1] {\vspace{12pt}\addtocounter{subsectionc}{1}
        \setcounter{subsubsectionc}{0}\noindent
        {\bf\thesectionc.\thesubsectionc. {\kern1pt \bfit #1}}\par\vspace{5pt}}
\renewcommand{\subsubsection}[1] {\vspace{12pt}\addtocounter{subsubsectionc}{1}
        \noindent{\tenrm\thesectionc.\thesubsectionc.\thesubsubsectionc.
        {\kern1pt \tenit #1}}\par\vspace{5pt}}
\newcommand{\nonumsection}[1] {\vspace{12pt}\noindent{\tenbf #1}
        \par\vspace{5pt}}

\newcounter{appendixc}
\newcounter{subappendixc}[appendixc]
\newcounter{subsubappendixc}[subappendixc]
\renewcommand{\thesubappendixc}{\Alph{appendixc}.\arabic{subappendixc}}
\renewcommand{\thesubsubappendixc}
        {\Alph{appendixc}.\arabic{subappendixc}.\arabic{subsubappendixc}}

\renewcommand{\appendix}[1] {\vspace{12pt}
        \refstepcounter{appendixc}
        \setcounter{figure}{0}
        \setcounter{table}{0}
        \setcounter{lemma}{0}
        \setcounter{theorem}{0}
        \setcounter{corollary}{0}
        \setcounter{definition}{0}
        \setcounter{equation}{0}
        \renewcommand{\thefigure}{\Alph{appendixc}.\arabic{figure}}
        \renewcommand{\thetable}{\Alph{appendixc}.\arabic{table}}
        \renewcommand{\theappendixc}{\Alph{appendixc}}
        \renewcommand{\thelemma}{\Alph{appendixc}.\arabic{lemma}}
        \renewcommand{\thetheorem}{\Alph{appendixc}.\arabic{theorem}}
        \renewcommand{\thedefinition}{\Alph{appendixc}.\arabic{definition}}
        \renewcommand{\thecorollary}{\Alph{appendixc}.\arabic{corollary}}
        \renewcommand{\theequation}{\Alph{appendixc}.\arabic{equation}}
        \noindent{\tenbf Appendix \theappendixc #1}\par\vspace{5pt}}
\newcommand{\subappendix}[1] {\vspace{12pt}
        \refstepcounter{subappendixc}
        \noindent{\bf Appendix \thesubappendixc. {\kern1pt \bfit #1}}
        \par\vspace{5pt}}
\newcommand{\subsubappendix}[1] {\vspace{12pt}
        \refstepcounter{subsubappendixc}
        \noindent{\rm Appendix \thesubsubappendixc. {\kern1pt \tenit #1}}
        \par\vspace{5pt}}

\newcommand{\textlineskip}{\baselineskip=13pt}
\newcommand{\smalllineskip}{\baselineskip=10pt}

\def\eightcirc{
\begin{picture}(0,0)
\put(4.4,1.8){\circle{6.5}}
\end{picture}}
\def\eightcopyright{\eightcirc\kern2.7pt\hbox{\eightrm c}}

\newcommand{\copyrightheading}[1]
        {\vspace*{-2.5cm}\smalllineskip{\flushleft
        {\footnotesize $\eightcopyright$\, World Scientific Publishing
         Company}\\
         }}

\def\abstracts#1#2#3{{
        \centering{\begin{minipage}{4.5in}\baselineskip=10pt\footnotesize
        \parindent=0pt #1\par
        \parindent=15pt #2\par
        \parindent=15pt #3
        \end{minipage}}\par}}

\newcommand{\bibit}{\nineit}
\newcommand{\bibbf}{\ninebf}
\renewenvironment{thebibliography}[1]
        {\frenchspacing
         \ninerm\baselineskip=11pt
         \begin{list}{\arabic{enumi}.}
        {\usecounter{enumi}\setlength{\parsep}{0pt}
         \setlength{\leftmargin 12.7pt}{\rightmargin 0pt} 
         \setlength{\itemsep}{0pt} \settowidth
        {\labelwidth}{#1.}\sloppy}}{\end{list}}

\newcounter{itemlistc}
\newcounter{romanlistc}
\newcounter{alphlistc}
\newcounter{arabiclistc}
\newenvironment{itemlist}
        {\setcounter{itemlistc}{0}
         \begin{list}{$\bullet$}
        {\usecounter{itemlistc}
         \setlength{\parsep}{0pt}
         \setlength{\itemsep}{0pt}}}{\end{list}}

\newenvironment{romanlist}
        {\setcounter{romanlistc}{0}
         \begin{list}{$($\roman{romanlistc}$)$}
        {\usecounter{romanlistc}
         \setlength{\parsep}{0pt}
         \setlength{\itemsep}{0pt}}}{\end{list}}

\newenvironment{alphlist}
        {\setcounter{alphlistc}{0}
         \begin{list}{$($\alph{alphlistc}$)$}
        {\usecounter{alphlistc}
         \setlength{\parsep}{0pt}
         \setlength{\itemsep}{0pt}}}{\end{list}}

\newcommand{\fcaption}[1]{
        \refstepcounter{figure}
        \setbox\@tempboxa = \hbox{\footnotesize Fig.~\thefigure. #1}
        \ifdim \wd\@tempboxa > 5in
           {\begin{center}
        \parbox{5in}{\footnotesize\smalllineskip Fig.~\thefigure. #1}
            \end{center}}
        \else
             {\begin{center}
             {\footnotesize Fig.~\thefigure. #1}
              \end{center}}
        \fi}

\newcommand{\tcaption}[1]{
        \refstepcounter{table}
        \setbox\@tempboxa = \hbox{\footnotesize Table~\thetable. #1}
        \ifdim \wd\@tempboxa > 5in
           {\begin{center}
        \parbox{5in}{\footnotesize\smalllineskip Table~\thetable. #1}
            \end{center}}
        \else
             {\begin{center}
             {\footnotesize Table~\thetable. #1}
              \end{center}}
        \fi}

\def\@citex[#1]#2{\if@filesw\immediate\write\@auxout
        {\string\citation{#2}}\fi
\def\@citea{}\@cite{\@for\@citeb:=#2\do
        {\@citea\def\@citea{,}\@ifundefined
        {b@\@citeb}{{\bf ?}\@warning
        {Citation `\@citeb' on page \thepage \space undefined}}
        {\csname b@\@citeb\endcsname}}}{#1}}

\newif\if@cghi
\def\cite{\@cghitrue\@ifnextchar [{\@tempswatrue
        \@citex}{\@tempswafalse\@citex[]}}
\def\citelow{\@cghifalse\@ifnextchar [{\@tempswatrue
        \@citex}{\@tempswafalse\@citex[]}}
\def\@cite#1#2{{$\null^{#1}$\if@tempswa\typeout
        {IJCGA warning: optional citation argument
        ignored: `#2'} \fi}}

\def\pmb#1{\setbox0=\hbox{#1}
        \kern-.025em\copy0\kern-\wd0
        \kern.05em\copy0\kern-\wd0
        \kern-.025em\raise.0433em\box0}

\def\fnm#1{$^{\mbox{\scriptsize #1}}$}
\def\fnt#1#2{\footnotetext{\kern-.3em
        {$^{\mbox{\scriptsize #1}}$}{#2}}}

\def\fpage#1{\begingroup
\voffset=.3in
\thispagestyle{empty}\begin{table}[b]\centerline{\footnotesize #1}
        \end{table}\endgroup}

\def\runninghead#1#2{\pagestyle{myheadings}
\markboth{{\protect\footnotesize\it{\quad #1}}\hfill}
{\hfill{\protect\footnotesize\it{#2\quad}}}}
\font\tenrm=cmr10
\font\tenit=cmti10
\font\tenbf=cmbx10
\font\bfit=cmbxti10 at 10pt
\font\ninerm=cmr9
\font\nineit=cmti9
\font\ninebf=cmbx9
\font\eightrm=cmr8

\def\qed{\hbox{${\vcenter{\vbox{                        
   \hrule height 0.4pt\hbox{\vrule width 0.4pt height 6pt
   \kern5pt\vrule width 0.4pt}\hrule height 0.4pt}}}$}}

\renewcommand{\thefootnote}{\fnsymbol{footnote}}        

\def\bsc{{\sc a\kern-6.4pt\sc a\kern-6.4pt\sc a}}       
\def\bflatex{\bf L\kern-.30em\raise.3ex\hbox{\bsc}\kern-.14em
T\kern-.1667em\lower.7ex\hbox{E}\kern-.125em X}

\begin{document}
%
%
%
%
\input{epsf.sty}
%
%
%
%
%
%
%
%
%
%
%
%
%
%
%
%
%
%
%
%
%
%
%
%
%
%
%
%
%
%
%
%
%
%
%
%
%
%
%
%
%
%
%
%
%
%
%
%
%
%
%
%
%
%
%
%
%
%
%
%
%
%
%
%
%
%
%
%
%
%
%
%
%

%
%
%
%

%

%
%
%
%
%
%
\def\axowidth{0.5 }
\def\axoscale{1.0 }
\def\axoxoff{0 }
\def\axoyoff{0 }
\def\axoxo{0 }
\def\axoyo{0 }
\def\firstcall{1}
\def\Gluon(#1,#2)(#3,#4)#5#6{
%
%
\put(\axoxoff,\axoyoff){
}
}
\def\Photon(#1,#2)(#3,#4)#5#6{
%
%
\put(\axoxoff,\axoyoff){
}
}
\def\ZigZag(#1,#2)(#3,#4)#5#6{
%
%
\put(\axoxoff,\axoyoff){
}
}
\def\PhotonArc(#1,#2)(#3,#4,#5)#6#7{
%
%
\put(\axoxoff,\axoyoff){
}
}
\def\GlueArc(#1,#2)(#3,#4,#5)#6#7{
%
%
\put(\axoxoff,\axoyoff){
}
}
\def\ArrowArc(#1,#2)(#3,#4,#5){
%
%
\put(\axoxoff,\axoyoff){
}
}
\def\LongArrowArc(#1,#2)(#3,#4,#5){
%
%
\put(\axoxoff,\axoyoff){
}
}
\def\DashArrowArc(#1,#2)(#3,#4,#5)#6{
%
%
\put(\axoxoff,\axoyoff){
}
}
\def\ArrowArcn(#1,#2)(#3,#4,#5){
%
%
\put(\axoxoff,\axoyoff){
}
}
\def\LongArrowArcn(#1,#2)(#3,#4,#5){
%
%
\put(\axoxoff,\axoyoff){
}
}
\def\DashArrowArcn(#1,#2)(#3,#4,#5)#6{
%
%
\put(\axoxoff,\axoyoff){
}
}
\def\ArrowLine(#1,#2)(#3,#4){
%
%
\put(\axoxoff,\axoyoff){
}
}
\def\LongArrow(#1,#2)(#3,#4){
%
%
\put(\axoxoff,\axoyoff){
}
}
\def\DashArrowLine(#1,#2)(#3,#4)#5{
%
%
\put(\axoxoff,\axoyoff){
}
}
\def\Line(#1,#2)(#3,#4){
%
%
\put(\axoxoff,\axoyoff){
}
}
\def\DashLine(#1,#2)(#3,#4)#5{
%
%
\put(\axoxoff,\axoyoff){
}
}
\def\CArc(#1,#2)(#3,#4,#5){
%
%
\put(\axoxoff,\axoyoff){
}
}
\def\DashCArc(#1,#2)(#3,#4,#5)#6{
%
%
\put(\axoxoff,\axoyoff){
}
}
\def\Vertex(#1,#2)#3{
%
%
\put(\axoxoff,\axoyoff){
}
}
\def\Text(#1,#2)[#3]#4{
%
%
\dimen0=\axoxoff \unitlength
\dimen1=\axoyoff \unitlength
\advance\dimen0 by #1 \unitlength
\advance\dimen1 by #2 \unitlength
\makeatletter
\@killglue\raise\dimen1\hbox to\z@{\kern\dimen0 \makebox(0,0)[#3]{#4}\hss}
\ignorespaces
\makeatother
}
\def\BCirc(#1,#2)#3{
%
%
\put(\axoxoff,\axoyoff){
}
}
\def\GCirc(#1,#2)#3#4{
%
%
\put(\axoxoff,\axoyoff){
}
}
\def\EBox(#1,#2)(#3,#4){
%
%
\put(\axoxoff,\axoyoff){
}
}
\def\BBox(#1,#2)(#3,#4){
%
%
\put(\axoxoff,\axoyoff){
}
}
\def\GBox(#1,#2)(#3,#4)#5{
%
%
\put(\axoxoff,\axoyoff){
}
}
\def\Boxc(#1,#2)(#3,#4){
%
%
\put(\axoxoff,\axoyoff){
}
}
\def\BBoxc(#1,#2)(#3,#4){
%
%
\put(\axoxoff,\axoyoff){
}
}
\def\GBoxc(#1,#2)(#3,#4)#5{
%
%
\put(\axoxoff,\axoyoff){
}
}
\def\SetWidth#1{\def\axowidth{#1 }}
\def\SetScale#1{\def\axoscale{#1 }}
\def\SetOffset(#1,#2){\def\axoxoff{#1 } \def\axoyoff{#2 }}
\def\SetScaledOffset(#1,#2){\def\axoxo{#1 } \def\axoyo{#2 }}
\def\pfont{Times-Roman }
\def\fsize{10 }
\def\SetPFont#1#2{\def\pfont{#1 } \def\fsize{#2 }}
%
%
\makeatletter
\def\fmode{4 }
\def\@l@{l} \def\@r@{r} \def\@t@{t} \def\@b@{b}
\def\mymodetest#1{\ifx#1\end \let\next=\relax \else {
\if#1\@r@\global\def\fmodeh{-3 }\fi
\if#1\@l@\global\def\fmodeh{3 }\fi
\if#1\@b@\global\def\fmodev{-1 }\fi
\if#1\@t@\global\def\fmodev{1 }\fi
} \let\next=\mymodetest\fi \next}
\makeatother
\def\PText(#1,#2)(#3)[#4]#5{
%
%
\def\fmodev{0 }
\def\fmodeh{0 }
\mymodetest#4\end
\put(\axoxoff,\axoyoff){\makebox(0,0)[]{\special{"/\pfont findfont \fsize
 scalefont setfont #1 \axoxo add #2 \axoyo add #3
\fmode \fmodev add \fmodeh add \fsize (#5) \axoscale ptext }}}
}
\def\GOval(#1,#2)(#3,#4)(#5)#6{
%
%
\put(\axoxoff,\axoyoff){
}
}
\def\Oval(#1,#2)(#3,#4)(#5){
%
%
\put(\axoxoff,\axoyoff){
}
}
\let\eind=]
\def\DashCurve#1#2{\put(\axoxoff,\axoyoff){
}}
\def\Curve#1{\put(\axoxoff,\axoyoff){
}}
\def\kromme(#1,#2)#3{#1 \axoxo add #2 \axoyo add \ifx #3\eind\else
\expandafter\kromme\fi#3}
\def\LogAxis(#1,#2)(#3,#4)(#5,#6,#7,#8){
%
%
\put(\axoxoff,\axoyoff){
}
}
\def\LinAxis(#1,#2)(#3,#4)(#5,#6,#7,#8,#9){
%
%
\put(\axoxoff,\axoyoff){
}
}
\input rotate.tex
\makeatletter
\def\rText(#1,#2)[#3][#4]#5{
%
%
\ifnum\firstcall=1\global\def\firstcall{0}\rText(-10000,#2)[#3][]{#5}\fi
\dimen2=\axoxoff \unitlength
\dimen3=\axoyoff \unitlength
\advance\dimen2 by #1 \unitlength
\advance\dimen3 by #2 \unitlength
\@killglue\raise\dimen3\hbox to \z@{\kern\dimen2
\makebox(0,0)[#3]{
\ifx#4l{\setbox3=\hbox{#5}\rotl{3}}\else{
\ifx#4r{\setbox3=\hbox{#5}\rotr{3}}\else{
\ifx#4u{\setbox3=\hbox{#5}\rotu{3}}\else{#5}\fi}\fi}\fi}\hss}
\ignorespaces
}
\makeatother
\def\BText(#1,#2)#3{
%
%
\put(\axoxoff,\axoyoff){
}
}
\def\GText(#1,#2)#3#4{
%
%
\put(\axoxoff,\axoyoff){
}
}
\def\B2Text(#1,#2)#3#4{
%
%
\put(\axoxoff,\axoyoff){
}
}
\def\G2Text(#1,#2)#3#4#5{
%
%
\put(\axoxoff,\axoyoff){
}
}

\runninghead{An Algorithm for Small Momentum
Expansion of Feynman Diagrams}
\normalsize\textlineskip
\thispagestyle{empty}
\setcounter{page}{1}

\copyrightheading{}                     

\vspace*{0.88truein}

\fpage{1}
\vspace{-1cm}
\hfill BI-TP-95/19\\
\mbox{}
\hfill hep-ph/9505277\\
\mbox{}
\hfill May 1995\\
\vspace{2cm}

\centerline{\bf AN ALGORITHM FOR SMALL MOMENTUM EXPANSION }
\centerline{\bf OF FEYNMAN DIAGRAMS}
\vspace*{0.37truein}
\centerline{\footnotesize O.\ V.\ TARASOV\footnote{
{}~~On leave from absence from Joint Institute  for Nuclear Research,
 141980, Dubna, Russia.
\\ \hspace*{3mm}
E-mail: oleg@physik.uni-bielefeld.de
\\ \hspace*{3mm}
Supported by Bundesministerium f\"ur Forschung und Technologie.
}}

\vspace*{0.015truein}
\centerline{\footnotesize\it
 Fakult\"at f\"ur Physik, Universit\"at Bielefeld,}
\baselineskip=10pt
\centerline{\footnotesize \it D-33615 Bielefeld 1, Germany}
\vspace*{10pt}

\vspace*{0.225truein}

\vspace*{0.21truein}
\abstracts{
An algorithm for obtaining the Taylor coefficients of an
expansion of Feynman diagrams is proposed.
It is based on  recurrence relations which
can be applied to the propagator   as well as  to the
 vertex  diagrams.
As an application,  several coefficients of the Taylor
series expansion for the two-loop  non-planar vertex  and two-loop
propagator diagrams are calculated.
The results of the numerical evaluation  of these diagrams using
conformal mapping and Pad\'e approximants are given.}{}{}

\vspace*{10pt}

\textlineskip                  
\vspace*{12pt}                 

\vspace*{1pt}\textlineskip      
\section{Introduction}    
\vspace*{-0.5pt}
\noindent


Recently a new method to calculate Feynman diagrams
was proposed\cite{FT1}. It is based on the small momentum
expansion\cite{DT}, conformal mapping and construction of Pad\'e
approximants from several terms in the Taylor series
of the diagram.
The method was successfully applied to the evaluation
of two- and three- loop diagrams\cite{FT2}.
As it was observed, a suitably accurate approximations
to the integrals can be obtained with 20-30 coefficients
in the Taylor series. The computation  of two-loop
vertex diagrams reveals
the necessity of an  efficient algorithm for the
expansion of the diagrams w.r.t. external momenta.

To outline the problem, let us shortly describe the
existing approach to the small momentum expansion.
At  present  the only method of  expansion
is based on  differentiation of the diagram
w.r.t. external momenta.
For  the propagator
type integrals  the prescription for the small momentum expansion
was formulated  in Ref.\cite{DT}.
Any coefficient in the Taylor series w.r.t. the external
momentum $q^2$ can be obtained
by applying $\Box = \frac{\partial}{\partial q_{\mu}}
        \frac{\partial}{\partial q^{\mu}}$ in an appropriate
power to the diagram and putting the external momentum
to zero.

The general expansion of a scalar 3-point function  $C(q_1, q_2)$
can be written as

\begin{equation}
\label{eq:exptri}
C(q_1, q_2) = \sum^\infty_{l,m,n=0} a_{lmn} (q^2_1)^l (q^2_2)^m
(q_1 q_2)^n = \sum^\infty_{L=0} \sum_{l+m+n=L} a_{lmn}
(q^2_1)^l (q^2_2)^m (q_1 q_2)^n.
\label{2.2}
\end{equation}

The coefficients $a_{lmn}$ are to be determined from a given diagram.
They can be obtained by applying the differential operators
$\Box_{ij} = \frac{\partial}{\partial p_{i\mu}}
\frac{\partial}{\partial p_j^\mu}$
several times to both sides of (\ref{eq:exptri}).
This procedure results in a system of linear equations for  $a_{lmn}$.
At a fixed $L$,
 one obtains several  systems of  $[L/2]+1$ equations.

Differential operators ($Df$'s),
projecting out the coefficients $a_{00n}$ in an arbitrary
space-time dimension $d$, are:
\begin{equation}
Df_{00n}=\sum^{[\frac{n}{2}]+1}_{i=1}
 \frac{(-4)^{1-i}\Gamma(d/2+n-i)\Gamma(d-1)}
 {2 \Gamma(i) \Gamma(n-2i+3) \Gamma(n+d-2) \Gamma(n+d/2)}(\Box_{12})^{n-2i+2}
 (\Box_{11}\Box_{22})^{i-1}.
\label{3.3}
\end{equation}
Applying  $Df_{00n}$ to
$C(q_1,q_2)$ and putting the external momenta equal to zero yields the
expansion coefficients $a_{00n}$.
For the coefficients $a_{l0n}$ the following
projection operator was obtained:
\begin{equation}
Df_{l0n}=\frac{\Gamma(\frac{d}{2}+n)} {\Gamma(l+1)
 \Gamma( \frac{d}{2}+l+n)} \left (
 \frac{\Box_{11}}{4} \right )^l Df_{00n}.
\end{equation}
The projection operator for arbitrary $a_{lmn}$ is
yet unknown.

An essential element of the above expansion is multiple
differentiation w.r.t. multidimensional vectors.
It turns out  that the computer implementation
of algorithms for the analytic calculation of the Taylor
coefficients based on differentiation is not very
effective. The computation of multiple sums with multiple
differentiations in an arbitrary space time dimension $d$,
even with the advanced possibilities
existing in FORM \cite{FORM}, leads to  significant
computational difficulties.

\vspace*{1pt}\textlineskip
\section{Recurrence relations}

 In the present paper we propose a new approach to this problem,
which we expect  to be more suitable
for the computer implementation.
The algorithm for the small
momentum expansion of scalar integrals can be formulated
as follows:

\begin{itemize}
\item
Firstly, the propagators with external momenta are to be
expanded as
$$
\frac{1}{(k_1^2-2k_1q_1+q_1^2-m_1^2)^j}=\frac{1}{(k_1^2+q_1^2-m_1^2)^j}
\sum_{l=0}^{\infty}
\frac{(l+j-1)!}{(j-1)! l!}
 \frac{(2k_1q_1)^l}{(k_1^2+q_1^2-m_1^2)^l}.
$$

This expansion  is valid in the whole integration
region since after the Wick rotation the expansion
parameter is always $\leq 1$.

\item
Secondly, by using recurrence relations the  external momenta
are to be factored out.

\item
Thirdly, the remaining factors should be expanded w.r.t. the
 external momenta $q_i^2$.
\item
Finally, using another kind of recurrence relations, bubble
integrals must be reduced to a set of master integrals
and trivial ones.
\end{itemize}

The proposed algorithm does not need any explicit
differentiation or solution of linear systems of equations.
All required
operations are efficiently  implemented in many computer
algebra systems. A key element  of the algorithm is the
recurrence relations allowing  one to factorize external and
integration momenta.
In the present paper such  recurrence relations will be
presented for the two-loop case. It's not difficult to get
them for  higher loop diagrams. For the three-loop case they
will be presented in a future publication.

 We shall now describe  a realization of the algorithm
for the two-loop propagator
and 3-point vertex diagrams.  At the one-loop level
the small momentum expansion
is more simple, and the algorithm will be evident
from the two-loop consideration. At the two- loop level
there are two
topologically different 3-point vertex diagrams  and
one propagator type topology (see $Fig.1$). We assume that the lines
in $Fig.1$  correspond to arbitrary
powers of the propagators.
\begin {figure} [htbp]
\begin{picture}(360,190)(-35,10)
\ArrowLine(10,100)(60,100)
\ArrowLine(10,150)(60,150)
\ArrowLine(10,100)(10,150)
\ArrowLine(60,150)(60,100)
\ArrowLine(60,100)(35,70)
\ArrowLine(35,70)(10,100)
\ArrowLine(35,50)(35,70)
\ArrowLine(10,150)(10,170)
\ArrowLine(60,170)(60,150)


\Text(-8,130)[]{$k_2+q_1$}
\Text(78,110)[]{$k_2+q_2$}
\Text(35,160)[]{$k_2$}
\Text(35,45)[]{$q_1-q_2$}
\Text(5,85)[]{$k_1+q_1$}
\Text(65,80)[]{$k_1+q_2$}
\Text(35,107)[]{$k_1-k_2$}
\Text(10,175)[]{$q_1$}
\Text(60,175)[]{$q_2$}

\Vertex(10,100){2}
\Vertex(60,100){2}
\Vertex(10,150){2}
\Vertex(60,150){2}
\Vertex(35,70){2}

\Line(110,100)(135,125)
\Line(110,150)(135,125)
\ArrowLine(135,125)(160,150)
\ArrowLine(135,125)(160,100)
\ArrowLine(110,100)(110,150)
\ArrowLine(160,150)(160,100)
\ArrowLine(135,70)(110,100)
\ArrowLine(160,100)(135,70)


\ArrowLine(135,50)(135,70)
\ArrowLine(110,150)(110,170)
\ArrowLine(160,170)(160,150)

\Vertex(110,100){2}
\Vertex(160,150){2}
\Vertex(135,70){2}
\Vertex(110,150){2}
\Vertex(160,100){2}

\Text(92,145)[]{$k_2+q_1$}
\Text(190,140)[]{$k_1-k_2+q_2$}
\Text(145,110)[]{$k_2$}
\Text(135,45)[]{$q_1-q_2$}
\Text(105,85)[]{$k_1+q_1$}
\Text(170,85)[]{$k_1+q_2$}
\Text(140,150)[]{$k_1-k_2$}
\Text(110,175)[]{$q_1$}
\Text(160,175)[]{$q_2$}

\ArrowLine(210,110)(225,110)
\ArrowLine(305,110)(320,110)
\ArrowLine(265,150)(265,70)

\ArrowArcn(265,110)(40,180,90)
\ArrowArcn(265,110)(40,90,0)
\ArrowArcn(265,110)(40,270,180)
\ArrowArcn(265,110)(40,0,-90)

\Vertex(225,110){2}
\Vertex(305,110){2}
\Vertex(265,150){2}
\Vertex(265,70){2}

\Text(240,128)[]{$k_1$}
\Text(308,130)[]{$k_2$}
\Text(285,110)[]{$k_1-k_2$}
\Text(230,74)[]{$k_1-q_1$}
\Text(310,74)[]{$k_2-q_1$}
\Text(145,15)[]{$Fig.1$}
\Text(218,100)[]{$q_1$}
\end{picture}
\end{figure}

 Both vertex diagrams, after expansion of propagators w.r.t.
scalar products $(k_iq_j)$, yield integrals of the type:

\begin{equation}
\int d^dk_1 ~d^dk_2~f(k_1,k_2,q_1^2,q_2^2)(k_1q_1)^{j_1}(k_1q_2)^{j_2}
 (k_2q_1)^{j_3}(k_2q_2)^{j_4}=v(j_1,j_2,j_3,j_4).
\label{v}
\end{equation}

These  satisfy the following recurrence
relation:
\begin{eqnarray}
&&(d+j_1+j_2+j_3+j_4-2)(d+j_1+j_3-3)v(j_1,j_2,j_3,j_4)=
 \nonumber \\
&& \nonumber \\
&& \left\{ (d+j_1+j_3+j_4-3) \left[ (j_1-1)
  k_1^2~q_1^2~{\bf 1^-}
 +j_2 k_1^2~ (q_1q_2)~{\bf  2^-}
+j_3~ (k_1k_2)~q_1^2~{ \bf 3^-} \right]
 \right. \nonumber \\
&& \nonumber \\
&&-j_3j_4 q_1^2 k_2^2 {\bf  2^+ 3^-4^-}
-j_4(j_4-1)k_2^2 (q_1q_2) {\bf 2^+ 4^-4^-}
- j_4(j_1-1) (k_1k_2) q_1^2~{\bf  1^- 2^+ 4^-}
 \nonumber \\
&& \nonumber \\
&&\left.+j_4(d+j_1-j_2+j_3+j_4-4)~(k_1k_2)~(q_1q_2)~{\bf 4^-}
 \right \}{\bf 1^-} \otimes v(j_1,j_2,j_3,j_4),
\label{recurrence}
\end{eqnarray}
where ${\bf 1}^{\pm}v(j_1,...)\equiv v(j_1\pm1,...)$ etc., and the
sign $\otimes$ means that the scalar products of the
integration momenta in  braces must
be considered under the integral
sign in $v$.
Relation (\ref{recurrence}) can be derived  from the following
tensor formula
\begin{eqnarray}
&& \int d^d k_1 d^dk_2 f(k_1,k_2,q_1^2,q_2^2) k_1^{\mu_1}...k_1^{\mu_N}
k_2^{\nu_1}...k_2^{\nu_L}=
 \nonumber \\
&&\nonumber \\
&&
{}~\frac{\Gamma(d/2) \Gamma(d-2)L!}{2^{(N+L)/2}}
P_{\nu} \sum_{p=0}^{[\frac{L}{2}]}
\frac{(L-2p+d/2-1)2^{-p}(L-2p)!}
 {p! \Gamma(L-p+d/2)\Gamma(L-2p+d-2)}
\nonumber \\
&& \nonumber \\
&&~~~\sum_{r=0}^{[\frac{L}{2}]-p}
\frac{\Gamma(L-2p-r+d/2-1)(-2)^{-r}}
{r! (L-2p-2r)! \Gamma(\frac{d+L+N}{2}-p-r)}
 g^{\nu_1 \nu_2}...g^{\nu_{2p+2r-1} \nu_{2p+2r}}
  \\
&& \nonumber \\
&&
{}~S^{[\mu_1,...,\mu_N,\nu_{2p+2r+1},...,\nu_{L}]}
 \int d^dk_1 d^dk_2 f(k_1,k_2,q_1^2,q_2^2)
 (k_1^2)^{\frac{N}{2}}
(k_2^2)^{\frac{L}{2}}
C^{(d/2-1)}_{L-2p} (\widehat{k_1k_2} ) \nonumber
\end{eqnarray}
after
contraction with $q_1,q_2$.
Here $C_p^{n}(x)$ are Gegenbauer polynomials,
$\widehat{k_1k_2}=(k_1 k_2)/ \sqrt{k_1^2k_2^2}$,
$S$ is the totally symmetric
sum of products of $g's$ and $P_{\nu}$
means symmetrization w.r.t. the indices $\nu$, i.e.
$P_{\nu}g^{\nu_1 \nu_2}S^{[...,\nu_3]}=
(g^{\nu_1\nu_2}S^{[...,\nu_3]}+g^{\nu_1 \nu_3}
S^{[...,\nu_2]}+g^{\nu_2 \nu_3}S^{[...,\nu_1]})/3$,
etc. The integral is equal to zero if $L+N$ is odd.
 In the above formula
$L<N$ is assumed  without loss of generality.

Expansion of the self-energy diagrams will give
integrals which correspond to  (\ref{v}) with
$j_1=j_3=0$.  In this case the recurrence relation
is very simple:
\begin{equation}
(d+j_2+j_4-2) v(0,j_2,0,j_4)=
q_2^2 \{
(j_2-1)k_1^2 {\bf 2^- } +j_4 (k_1k_2)
{\bf 4^-}\}{\bf 2^-}
\otimes v(0,j_2,0,j_4).
\label{recprop}
\end{equation}

In Ref.\cite{DST} an explicit formula for $v(0,j_2,0,j_4)$
in terms of  a onefold
sum is given. The computer implementation of that formula,
however,
is less effective than the application of the recurrence
relation (\ref{recprop}). Substitution of the sum blows up
expressions, producing too  many similar terms at once.

 A recurrence relation for the one-loop vertex diagrams
is also a special case of (\ref{recurrence}) with $j_3=j_4=0$:
\begin{equation}
(d+j_1+j_2-2) v(j_1,j_2,0,0)= k_1^2 [(j_1-1) q_1^2 {\bf 1^-}
+j_2 ~(q_1q_2) ~{\bf 2}^-] {\bf 1}^- \otimes v(j_1,j_2,0,0).
\end{equation}
Here it is assumed that the integration w.r.t. $k_2$ in
$v$ is omitted.

The recurrence relation (\ref{recurrence}) can also  be  used
for  evaluating  anomalous dimensions
of the moments of structure functions  in deep inelastic scattering
in the MOM renormalization scheme. In this case, one needs
to calculate propagator type integrals with vertex insertions
containing some additional vector $\xi$. The small momentum
expansion of these diagrams will produce integrals of the type

\begin{equation}
\int d^dk_1 d^dk_2 f(k_1,k_2,q_1^2) (k_1q_1)^{j_1} (k_1 \xi)^{j_2}
 (k_2q_1)^{j_3},
\end{equation}
again corresponding to the special case of (\ref{recurrence})
with $j_4 =0$:
\begin{eqnarray}
&&(d+j_1+j_2+j_3-2)v(j_1,j_2,j_3,0)= \\
&&~~~~~~~~~~~[(j_1-1)k_1^2 q_1^2 {\bf 1}^-
 +j_2 k_1^2 ( \xi q_1) {\bf 2}^- + j_3 (k_1 k_2) q_1^2 {\bf 3}^-
 ]{\bf 1}^- \otimes v(j_1,j_2,j_3,0).
\nonumber
\end{eqnarray}

After some modifications, the  recurrence relation
(\ref{recurrence})
can be used for the calculation of the Taylor coefficients
in the expansion of  diagrams in axial
gauges.

 By using   relation (\ref{recurrence})
and relations derived  from it by exchanging
$q_1 \leftrightarrow q_2$ and  $k_1 \leftrightarrow k_2$,
the  exponents  of the scalar products can be reduced
to zero.
One should choose the scalar product with the smallest
exponent, apply the recurrence relation until this
exponent is reduced to zero
and repeat this procedure  until only one
scalar product in some power remains. For these integrals
the following explicit formula can then  be used:
\begin{eqnarray}
&&\int f(k_1,k_2,q_1^2,q_2^2)~(k_1 q_1)^{2j_1} d^dk_1
d^dk_2  \\
&&~~~~~~~~~~~~~~~~~~~~~~~~ =\frac{(2j_1)!}{(d/2)_{j_1}}
 \left( \frac{q_1^2}{4} \right )^{j_1}
\int f(k_1,k_2,q_1^2,q_2^2)(k_1^2)^{j_1} d^dk_1 d^dk_2.
\nonumber
\end{eqnarray}

After that we obtain bubble-like integrals depending only
on the external momenta squared:
\begin{equation}
\int d^dk_1 ~d^dk_2~f(k_1,k_2,q_1^2,q_2^2)~(k_1^2)^{j_1}(k_1k_2)^{j_2}
 (k_2^2)^{j_3}.
\label{bubbles}
\end{equation}
The most optimal way to evaluate these integrals depends
on the mass values. If some masses are the same,
or one  mass is zero, then
it is worthwhile to expand the integrand w.r.t. $q_1^2,q_2^2$
and after that to calculate the bubble integrals using
recurrence relations
given in Refs.\cite{DT}$^,$\cite{FT3}. If the masses are different
it will be more efficient to evaluate the
integrals (\ref{bubbles}) using the recurrence relations
before the expansion in $q_1^2,q_2^2$ is performed.
In this case the presence of  $q_1^2,q_2^2$ in the denominators
can be considered as a  mass shift, and therefore, the same
recurrence relations can be applied.

\section{Applications}
To check our algorithm, we repeated the calculation
of the diagram that occurs in the process
$H \rightarrow \gamma \gamma$ \cite{FT1}. In comparison
with the old algorithm based on differentiation,
the execution time reduced by a factor of about 20.
For example, with the old algorithm the evaluation
of the 28-th coefficient
in the Taylor series for this diagram took 5
hours CPU time on the DEC 3000.
With the new algorithm, running FORM \cite{FORM}
on the same computer, it took 15 min  to obtain
the result. Further improvement in the efficiency of the computation
can be achieved if the algorithm can be implemented
in a multiple precision FORTRAN program \cite{Bailey}.
In this case, as it was observed
in \cite{FT3}$^,$\cite{JF} , the numerical approach
should substantially more efficient than the analytical one.

The method proposed here was  used for the evaluation
of the Taylor coefficients of the nonplanar scalar vertex
and propagator diagrams. In both cases
all internal lines were considered to be  massive. For simplicity
the same mass was taken and
 $q_1^2=q_2^2=0$ was chosen, so
that the diagram  depended on just one variable $q^2=-2q_1q_2$.
The evaluation of the non-planar diagram by numerical methods
was complicated due to the  integrable singularities
\cite{jap}. It is interesting to investigate
how the method proposed in\cite{FT1} works in this
case.
The diagram has a cut for $q^2 \geq 4m^2$. After the conformal
mapping  we find a good convergence of the sequence of Pad\'e
approximants derived from several first
Taylor coefficients. Table 1 demonstrates the quality
of the convergence and contains values (up to a factor
$10^{-9}/(16\pi^2)^2$) for the diagram on the
cut. Errors were estimated by comparison of the $[8/8]$ and
$[9/9]$ Pad\'e approximants.
{\scriptsize
\hsize=11in\vsize=8in
\begin{table}[htb]
\caption{Results for [9/9] on the cut $(q^2 > 4m^2_t),m= m_t = 150$ GeV}
\medskip
\def\.{&.}\def\pl{&$\pm$}
\halign{\strut\vrule~\hfil#&#~\vrule
&~\hfil#&#
&~\hfil#&#~\vrule &~\hfil#&# &~\hfil#&#~\vrule  &~\hfil#&#
&~\hfil#&#~\vrule
\cr
\noalign{\hrule}
& $q^2/m^2_t$ && {\it nonplanar } : && errors &&
 {\it propagator} :errors &&&&  {\it nonplanar }:&&results
\cr
& && {\it vertex} &&&&&&&&{\it vertex} && \cr

&&& Re && Im && Re && Im && Re && Im \cr
\noalign{\hrule}
4&.00
&&$\Pp188$& &0.         & &$\Pp577$ & &0.       & 0&.733120
& 0&.         \cr
4&.01
&&$\Pp779$& &$\Pp237$ & &$\Pp808$ & &$\Pp446$ & 0&.73056733
&--0&.0523635 \cr
4&.05
&&$\Pp178$& &$\Pp718$ & &$\Pp647$ & &$\Pp117$ & 0&.7204535
&--0&.1160966 \cr
4&.50
&&$\Pp139$& &$\Pp127$ & &$\Pp196$    & &$\Pp527$ & 0&.61644824
&--0&.3349475 \cr
5&.00
&&$\Pp268$& &$\Pp277$& &$\Pp526$  & &$\Pp226$ & 0&.5184444
&--0&.430997  \cr
8&.00
&&$\Pp686$& &$\Pp495$& &$\Pp464$  & &$\Pp735$ &  0&.14555
&--0&.5460    \cr
9&.0
&&$\Pp844$& &$\Pp194$& &$\Pp624$  & &$\Pp204$ & 0&.0613
&--0&.539     \cr
10&.0
&&$\Pp913$& &$\Pp324$& &$\Pp254$  & &$\Pp384$ & 0&.018
&--0&.516     \cr
20&.0
&&$\Pp805$& &$\Pp384$& &$\Pp365$  & &$\Pp154$ &--0&.2047
&--0&.1876    \cr
40&.0
&&$\Pp314$& &$\Pp143$& &$\Pp695$  & &$\Pp204$ &--0&.1259
&--0&.0225   \cr
100&.0
&&$\Pp324$& &$\Pp333$& &$\Pp214$  & &$\Pp344$ &--0&.0382
& 0&.0152 \cr
200&.0
&&$\Pp313$& &$\Pp403$& &$\Pp185$  & &$\Pp533$ &--0&.0125
& 0&.0105 \cr
400&.0
&&$\Pp132$& &$\Pp833$& &$\Pp964$  & &$\Pp383$ &--0&.0036
& 0&.00511 \cr
\noalign{\hrule}}
\label{pade}
\end{table}
}
With 18 coefficients, as one can see
from Table 1,  the accuracy of
calculations for $q^2<400 m^2$ is better than
$1 \%$. The convergence of the approximants below the cut
is substantially better than on the cut.
In Table 1 we also give  results on the cut for the propagator
diagram. The precision of the results for the propagator  diagram
is practically the same as for the vertex one.
This fact can be an indication  that
for the diagrams with arbitrary scalar products of the
external momenta, taking in (1)  $L=18-20$, we can achieve
a similar accuracy.

\textheight=7.8truein
\setcounter{footnote}{0}
\renewcommand{\thefootnote}{\alph{footnote}}


\nonumsection{Acknowledgments}
I am  grateful to L.Avdeev, J.Fleischer and V.Smirnov
for useful discussions and a careful reading of the manuscript
and to A.Davydychev  for useful discussions.

\vspace{-2mm}

\nonumsection{References}

\end{document}

\nonumsection{References}
\noindent
References are to be listed in the order cited in the text. Use
the style shown in the following examples. For journal names,
use the standard abbreviations. Typeset references in 9 pt Times
roman.


Contributions to the {\it International Journal of Modern
Physics C} will be reproduced by photographing the author's
submitted typeset manuscript. It is therefore essential that the
manuscript be in its final form, and of good appearance because
it will be printed directly without any editing. The manuscript
should also be clean and unfolded. The copy should be evenly
printed on a high resolution printer (300 dots/inch or higher).
If typographical errors cannot be avoided, use cut and paste
methods to correct them. Smudged copy, pencil or ink text
corrections will not be accepted. Do not use cellophane or
transparent tape on the surface as this interferes with the
picture taken by the publisher's camera.
\pagebreak

\textheight=7.8truein
\setcounter{footnote}{0}
\renewcommand{\thefootnote}{\alph{footnote}}

\section{The Main Text}
\noindent
Contributions are to be in English. Authors are encouraged to
have their contribution checked for grammar. American spelling
should be used. Abbreviations are allowed but should be spelt
out in full when first used. Integers ten and below are to be
spelt out. Italicize foreign language phrases (e.g.~Latin,
French).

The text is to be typeset in 10 pt Times roman, single spaced
with baselineskip of 13 pt. Text area (excluding running title)
is 5 inches (30 picas) across and 7.8 inches (47 picas) deep.
Final pagination and insertion of running titles will be done by
the publisher. Number each page of the manuscript lightly at the
bottom with a blue pencil. Reading copies of the paper can be
numbered using any legible means (typewritten or handwritten).

\section{Major Headings}
\noindent
Major headings should be typeset in boldface with the first
letter of important words capitalized.

\subsection{Sub-headings}
\noindent
Sub-headings should be typeset in boldface italic and capitalize
the first letter of the first word only. Section number to be in
boldface roman.

\subsubsection{Sub-subheadings}
\noindent
Typeset sub-subheadings in medium face italic and capitalize the
first letter of the first word only. Section numbers to be in
roman.

\subsection{Numbering and spacing}
\noindent
Sections, sub-sections and sub-subsections are numbered in
Arabic.  Use double spacing before all section headings, and
single spacing after section headings. Flush left all paragraphs
that follow after section headings.

\subsection{Lists of items}
\noindent
Lists may be laid out with each item marked by a dot:
\begin{itemlist}
 \item item one,
 \item item two.
\end{itemlist}
Items may also be numbered in lowercase roman numerals:
\begin{romanlist}
 \item item one
 \item item two
        \begin{alphlist}
        \item Lists within lists can be numbered with lowercase
              roman letters,
        \item second item.
        \end{alphlist}
\end{romanlist}
\newpage

\section{Equations}
\noindent
Displayed equations should be numbered consecutively in each
section, with the number set flush right and enclosed in
parentheses
\begin{equation}
\mu(n, t) = {\sum^\infty_{i=1} 1(d_i < t, N(d_i) = n) \over
\int^t_{\sigma=0} 1(N(\sigma) = n)d\sigma}\,. \label{this}
\end{equation}

Equations should be referred to in abbreviated form,
e.g.~``Eq.~(\ref{this})'' or ``(\ref{that})''. In multiple-line
equations, the number should be given on the last line.

Displayed equations are to be centered on the page width.
Standard English letters like x are to appear as $x$
(italicized) in the text if they are used as mathematical
symbols. Punctuation marks are used at the end of equations as
if they appeared directly in the text.

\vspace*{12pt}
\noindent
{\bf Theorem~1:} Theorems, lemmas, etc. are to be numbered
consecutively in the paper. Use double spacing before and after
theorems, lemmas, etc.

\vspace*{12pt}
\noindent
{\bf Proof:} Proofs should end with \qed\,.

\section{Illustrations and Photographs}
\noindent
Figures are to be inserted in the text nearest their first
reference.  Original india ink drawings of glossy prints are
preferred. Please send one set of originals with copies. If the
author requires the publisher to reduce the figures, ensure that
the figures (including letterings and numbers) are large enough
to be clearly seen after reduction. If photographs are to be
used, only black and white ones are acceptable.

\begin{figure}[htbp]
\vspace*{13pt}
\centerline{\vbox{\hrule width 5cm height0.001pt}}
\vspace*{1.4truein}             
\centerline{\vbox{\hrule width 5cm height0.001pt}}
\vspace*{13pt}
\fcaption{Labeled tree {\footnotesize\it T}.}
\end{figure}

Figures are to be sequentially numbered in Arabic numerals. The
caption must be placed below the figure. Typeset in 8 pt Times
roman with baselineskip of 10 pt. Use double spacing between a
caption and the text that follows immediately.

Previously published material must be accompanied by written
permission from the author and publisher.
\pagebreak

\section{Tables}
\noindent
Tables should be inserted in the text as close to the point of
reference as possible. Some space should be left above and below
the table.

Tables should be numbered sequentially in the text in Arabic
numerals. Captions are to be centralized above the tables.
Typeset tables and captions in 8 pt Times roman with
baselineskip of 10 pt.

\begin{table}[htbp]
\tcaption{Number of tests for WFF triple NA = 5, or NA = 8.}
\centerline{\footnotesize NP}
\centerline{\footnotesize\smalllineskip
\begin{tabular}{l c c c c c}\\
\hline
{} &{} &3 &4 &8 &10\\
\hline
{} &\phantom03 &1200 &2000 &\phantom02500 &\phantom03000\\
NC &\phantom05 &2000 &2200 &\phantom02700 &\phantom03400\\
{} &\phantom08 &2500 &2700 &16000 &22000\\
{} &10 &3000 &3400 &22000 &28000\\
\hline\\
\end{tabular}}
\end{table}

If tables need to extend over to a second page, the continuation
of the table should be preceded by a caption, e.g.~``{\it Table
2.} $(${\it Continued}$)$''

\section{References}
\noindent
References in the text are to be numbered consecutively in
Arabic numerals, in the order of first appearance. They are to
be typed in superscripts after punctuation marks,
e.g.~``$\ldots$ in the statement.$^5$''.

\section{Footnotes}
\noindent
Footnotes should be numbered sequentially in superscript
lowercase roman letters.\fnm{a}\fnt{a}{Footnotes should be
typeset in 8 pt Times roman at the bottom of the page.}

\nonumsection{Acknowledgements}
\noindent
This section should come before the References. Funding
information may also be included here.

\nonumsection{References}
\noindent
References are to be listed in the order cited in the text. Use
the style shown in the following examples. For journal names,
use the standard abbreviations. Typeset references in 9 pt Times
roman.

\appendix

\noindent
Appendices should be used only when absolutely necessary. They
should come after the References. If there is more than one
appendix, number them alphabetically. Number displayed equations
occurring in the Appendix in this way, e.g.~(\ref{that}), (A.2),
etc.
\begin{equation}
\mu(n, t) = {\sum^\infty_{i=1} 1(d_i < t, N(d_i) = n) \over
\int^t_{\sigma=0} 1(N(\sigma) = n)d\sigma}\,. \label{that}
\end{equation}
\end{document}